\documentclass[aps,prl,twocolumn,showpacs,superscriptaddress]{revtex4-1}
\usepackage{graphicx}  
\usepackage{dcolumn}   
\usepackage{bm}        
\usepackage{amssymb}   
\usepackage{amsfonts}
\usepackage{verbatim}

\usepackage{amsmath}
\usepackage{amsfonts}
\usepackage{ulem}
\usepackage{color}

\newcommand{\ket}[1]{|#1\rangle}

\def\lsim{\mathrel{\rlap{\lower4pt\hbox{\hskip1pt$\sim$}}
    \raise1pt\hbox{$<$}}}                
\def\gsim{\mathrel{\rlap{\lower4pt\hbox{\hskip1pt$\sim$}}
    \raise1pt\hbox{$>$}}}                

\begin{document}
\normalem

\title{Microwave Photon Counter Based on Josephson Junctions}

\author{Y.-F. Chen}
\altaffiliation[Present address: ]{Department of Physics, National Central
University, Jung-Li 32001, Taiwan.}
\author{D. Hover}
\author{S. Sendelbach}
\author{L. Maurer}
\affiliation{Department of Physics, University of
Wisconsin, Madison, Wisconsin 53706, USA}
\author{S. T. Merkel}
\altaffiliation[Present address: ]{Department of Theoretical Physics, Saarland University, Saarbrucken, Germany.}
\author{E. J. Pritchett}
\altaffiliation[Present address: ]{Department of Theoretical Physics, Saarland University, Saarbrucken, Germany.}
\author{F. K. Wilhelm}
\altaffiliation[Present address: ]{Department of Theoretical Physics, Saarland University, Saarbrucken, Germany.}
\affiliation{Institute for Quantum Computing and Department of Physics and Astronomy, University of Waterloo, Waterloo, Canada}
\author{R. McDermott}\email[Electronic address: ]{rfmcdermott@wisc.edu}
\affiliation{Department of Physics, University of
Wisconsin, Madison, Wisconsin 53706, USA}

\date{\today}

\begin{abstract}
We describe a microwave photon counter based on the current-biased Josephson junction. The junction is tuned to absorb single microwave photons from the incident field, after which it tunnels into a classically observable voltage state. Using two such detectors, we have performed a microwave version of the Hanbury Brown and Twiss experiment at 4 GHz and demonstrated a clear signature of photon bunching for a thermal source. The design is readily scalable to tens of parallelized junctions, a configuration that would allow number-resolved counting of microwave photons.
\end{abstract}

\maketitle

The strong interaction of superconducting integrated circuits with microwave photons forms the basis of circuit quantum electrodynamics (cQED), an attractive paradigm for scalable quantum computing and a test bed for quantum optics in the strong coupling regime \cite{Blais04,Wallraff04,Chiorescu04,Johansson06,Hofheinz08,Hofheinz09}. While single-photon detectors are an important component of the standard quantum optics toolbox \cite{Walls08}, they have not played a role in cQED, as there are no microwave photon counters currently available. In the search for a microwave component that will emulate the nonlinearity of traditional optical photon counters, it is natural to consider the Josephson junction -- a nonlinear, nondissipative superconducting circuit element with tunable microwave-frequency transitions. There has been recent interest in the development of on-chip detectors to probe the microwave radiation emitted by mesoscopic quantum circuits \cite{Romero09,Romero09b,Aguado00,Beenakker01,Deblock03,Onac06,Gustavsson07}, including proposals to utilize Josephson circuits as photon counters in the context of cQED \cite{Romero09,Romero09b}. This study represents the first experimental measurement of the coincidence counting statistics of microwave photons. We exploit microwave-induced transitions between discrete energy levels in a Josephson junction to realize an efficient, versatile microwave detector. Moreover, we use a parallelized detector circuit to probe the counting statistics of both coherent and thermal microwave sources.

\begin{figure}[t]
\includegraphics[width=.48\textwidth]{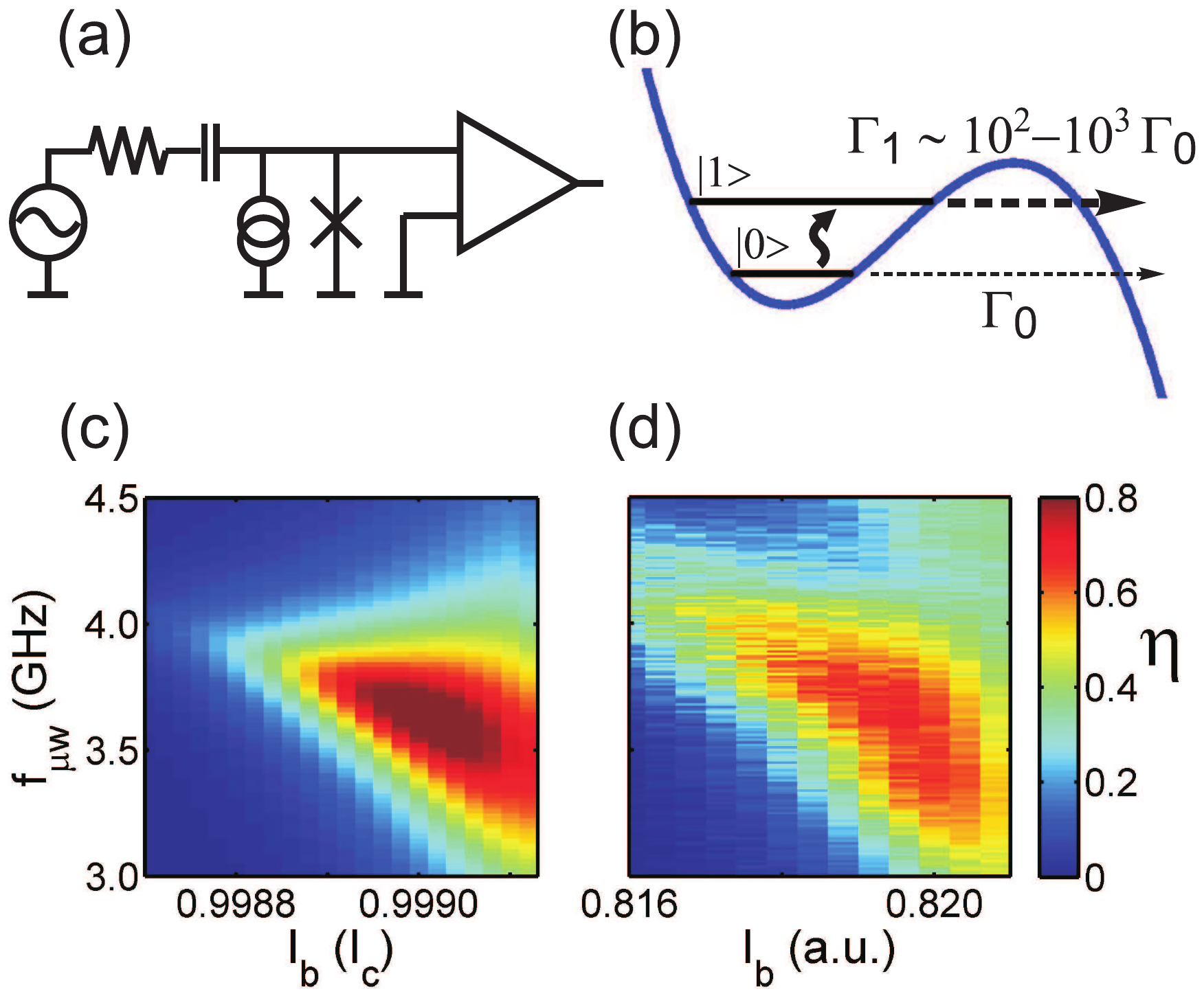}
\vspace*{-0.0in} \caption{Microwave photon detector based on a Josephson junction. (a) Schematic diagram of the circuit. The junction is biased with a dc current, and microwaves are coupled to the junction via an on-chip capacitor. The voltage across the junction is read out by a room temperature preamplifier and comparator. (b) Junction potential energy landscape. The junction is initialized in the $\ket{0}$ state. An incident photon induces a transition to the $\ket{1}$ state, which rapidly tunnels to the continuum. (c) Calculated detection efficiency $\eta$ as a function of applied microwave frequency and current bias. The simulation parameters are $I_0$ = 210 $\mu$A, $C$ = 45 pF, $T_1$ = 3 ns, $\omega_{\rm R}/2 \pi$ = 200 MHz, and $t_{\rm det}$ = 5 ns, chosen to match the parameters of our measured device. (d) Experimentally measured detection efficiency. The efficiency shows a maximum of 0.7. The detection bandwidth is of order the Rabi frequency associated with the microwave drive.}
\label{fig:figure1}\end{figure}

A circuit diagram of the single-channel detector is shown in Fig.~\ref{fig:figure1}(a). The Josephson junction is current biased near the critical current $I_0$. The junction potential energy is given by $U(\delta)=-\frac{I_0 \Phi_0}{2 \pi} \cos \delta - I_{\rm b} \delta$, where $I_{\rm b}$ is the bias current, $\delta$ is the superconducting phase difference across the junction, and $\Phi_0 \equiv h/2e$ is the magnetic flux quantum. For bias currents approaching the critical current, the local minima of the potential (characterized by barrier height $\Delta U$ and plasma frequency $\omega_{\rm p}$) accommodate a handful of discrete energy levels. Resonant microwaves induce coherent oscillations between the ground state $\ket{0}$ and the excited state $\ket{1}$ of the junction (Fig.~\ref{fig:figure1}(b)). Moreover, each of these states can tunnel to the continuum at a rate $\Gamma_{0,1}$. Due to the exponential dependence of tunneling on barrier height, $\Gamma_1$ is 2-3 orders of magnitude larger than $\Gamma_0$.

\begin{figure}[t]
\includegraphics[width=.48\textwidth]{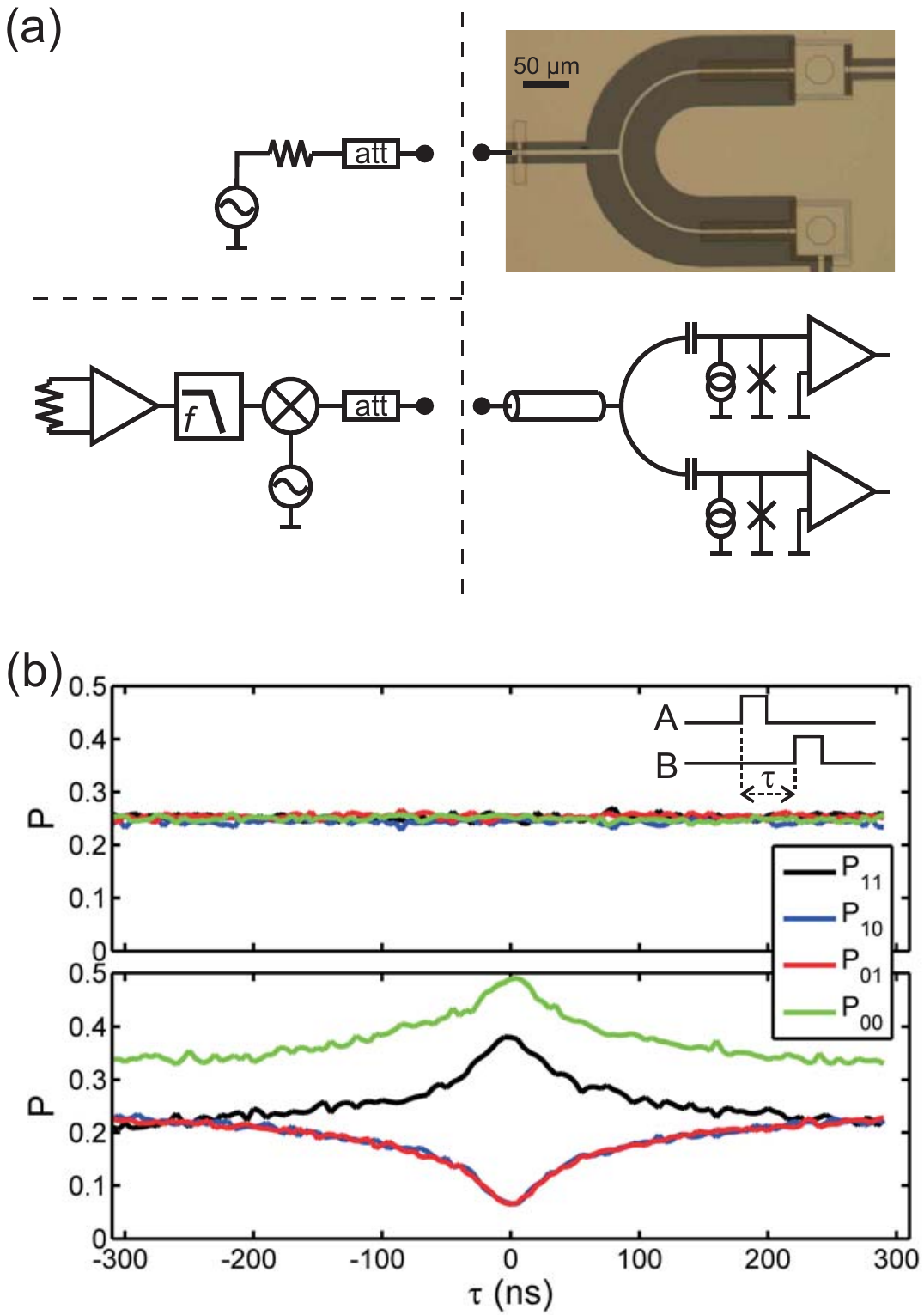}
\vspace*{-0.0in} \caption{(a) Experimental setup of microwave-frequency version of the HBT experiment. Coherent microwaves (\textit{upper left panel}) are derived from a microwave generator whose output is coupled to the detector via a transmission line, with suitable attenuation. The thermal source (\textit{lower left panel}) is realized by amplifying, filtering, and upconverting the Johnson noise of a room temperature resistor. A two-junction circuit (\textit{right panel}) is used to probe the statistical nature of the microwave photon source; each junction has its own bias and readout. (b) Conditional switching probabilities $P_{00}$, $P_{10}$, $P_{01}$, and $P_{11}$ \textit{versus} relative measurement delay $\tau$ for coherent microwaves (\textit{upper panel}) and thermal microwaves (\textit{lower panel}). Inset shows the timing diagram of the measurement pulses applied to the two junctions.}
\label{fig:figure2}\end{figure}

Our experimental protocol involves pulsing the junction bias for a finite interval (of order 10 ns) during which the transition frequency $\omega_{10}$ is near resonant with the incident photons. Absorption of a photon promotes the junction to the $\ket{1}$ state, which tunnels rapidly to the continuum, producing a large voltage pulse and a measurable ``click". We have calculated the time evolution of the junction density matrix in the presence of a classical coherent drive. We take into account $T_1$ relaxation and tunneling to the continuum using standard open quantum systems tools \cite{Breuer02}. To quantify the performance of the detector, we introduce the efficiency parameter $\eta \equiv P_{\rm bright}(1-P_{\rm dark})$, where $P_{\rm bright}$ is the probability of a microwave-induced transition to the continuum during the measurement interval and $P_{\rm dark}$ is the probability of a spurious dark count due to quantum tunneling. We find that $\eta$ is maximized for a bias point corresponding to $\Delta U/\hbar \omega_{\rm p} \sim 2$ and for a measurement interval that is roughly equal to the Rabi period of the coherent drive; the detection bandwidth is of the order of the Rabi frequency $\omega_{\rm R}/2 \pi$. We note that for $\omega_{\rm R}$ comparable to vacuum Rabi frequencies achieved in cQED experiments ($\omega_{\rm R}/2 \pi \sim$ 100 MHz), detection efficiencies in excess of 0.7 are attained with junctions having extremely modest coherence times of order several ns. 

It is instructive to compare our Josephson photon counter to other Josephson junction-based detectors. Superconducting tunnel junction detectors are biased in the voltage state and incident photons (typically at near infrared frequencies and beyond) break Cooper pairs to induce an enhanced subgap current \cite{Gatti04}; in contrast, our junctions are biased in the supercurrent state, and incident microwave photons induce a transition to the voltage state. In other recent work, the bifurcation of a driven Josephson junction has been exploited to realize sensitive threshold detectors \cite{Siddiqi04} or amplifiers for field quadratures \cite{Vijay11}. In contrast, our detector is strictly sensitive to the photon number: it registers a click whenever a nonzero number of photons is present, irrespective of that number.

We have fabricated a double photon counter sample in which two Josephson junctions (labeled $A$ and $B$) with independent biasing and readout are driven by a common microwave source. The junctions were made in a standard Al-AlO$_{\rm x}$-Al technology, with areas of 1000~$\mu$m$^2$. The detectors are operated at 37~mK in a dilution refrigerator in order to minimize the thermal population of the excited state. We first characterized each junction separately by performing spectroscopy and Rabi experiments (not shown). The junction $T_1$ times are approximately 3 ns, compatible with the $RC$ time expected for the $\sim$ 45 pF junctions, where $R$ is an environmental impedance of order 100 $\Omega$. In Fig.~\ref{fig:figure1}(c) we have used the measured junction parameters to simulate efficiency $\eta$ for one of these junctions as a function of coherent drive frequency and detector bias for $\omega_{\rm R}/2 \pi$~=~200~MHz. The experimental data shown in Fig.~\ref{fig:figure1}(d) are in good qualitative agreement with the model. For a bias point corresponding to a resonance frequency $\omega_{10}/2 \pi = 3.8$~GHz, the peak detection efficiency is 0.7; the efficiency exceeds 0.6 over a band larger than 450 MHz. The appreciable detection efficiency and bandwidth of these detectors are well-suited to cQED experiments involving high quality on-chip cavities.

Next, we examined correlations between the two detectors for different types of microwave source. The microwaves were coupled to the junctions via an on-chip power divider with coupling capacitors at each output port (Fig.~\ref{fig:figure2}(a)). With microwaves applied, we sequentially pulsed the two junctions into the active state for 5~ns, varying the relative delay $\tau$ of the two measurement pulses. The experiments were repeated $10^4 - 10^5$ times, and we recorded the joint probabilities $P_{00},P_{01},P_{10},P_{11}$ for the two detectors to remain in the supercurrent state (outcome ``$\,$0$\,$") or switch to the voltage state (outcome ``$\,$1$\,$").

In the first experiment, we applied 4 GHz coherent photons derived from a commercial microwave generator (Fig.~\ref{fig:figure2}(a), upper left panel). The microwaves from a commercial generator constitute a classical electromagnetic wave with well-defined phase and amplitude, and provide a good approximation to coherent radiation \cite{Hofheinz08, Gabelli04, Schuster07}. In the upper panel of Fig.~\ref{fig:figure2}(b) we plot the joint switching probabilities versus $\tau$. We observe no correlations in the switching of the two junctions and see that $P_{11}$ = $P_1^{\rm A} P_1^{\rm B}$, as expected if absorption of a photon is a homogeneous Poisson process. In the second experiment, we applied thermal microwaves to the double-junction detector. The thermal source was realized by amplifying and filtering the Johnson noise from a resistor at room temperature and upconverting the noise to our 4 GHz detection frequency using a microwave mixer (Fig.~\ref{fig:figure2}(a), lower left panel). The results of irradiation with the thermal source are shown in the lower panel of Fig.~\ref{fig:figure2}(b). For short relative delay, we observe a clear enhancement of $P_{00}$ and $P_{11}$, accompanied by a suppression of $P_{01}$ and $P_{10}$. The switching probabilities of the individual junctions $P_1^{\rm A} = P_{11}+P_{10}$ and $P_1^{\rm B}=P_{11}+P_{01}$ do not depend on the relative delay of the two measurements.

These experiments constitute a microwave-frequency realization of the classic Hanbury Brown and Twiss (HBT) experiment \cite{Hanbury56}, in which temporal correlations in the intensity fluctuations at two detectors yield information about the statistical nature of the photon source. We define an enhancement factor $R_{11}(\tau) \equiv  P_{11}(\tau)/P_{11}(\tau \rightarrow \infty)= P_{11}(\tau)/P_1^{\rm A} P_1^{\rm B}$. In the standard quantum optics treatment of photon counting where the detector is understood to interact weakly with the photon field, one can show that $R_{11}(\tau)$ is equivalent to the second-order quantum coherence of the field $g^{(2)}(\tau)$ \cite{Walls08}. For a coherent state, the photon arrival times are uncorrelated and one finds that $g^{(2)}=1$ for all relative delays, while for a thermal state one finds that $g^{(2)}$ attains a value of 2 for $\tau=0$ and decays to 1 on a timescale set by the coherence time of the source. The factor of 2 enhancement of $g^{(2)}$ is a manifestation of the Bose statistics of the thermal source: one observes enhanced probability for measuring two (or zero) photons relative to the result obtained for a homogeneous Poisson process within a timescale set by the characteristic fluctuations of the source. The phenomenon of photon bunching has been well established at optical frequencies by numerous experiments. Clear evidence of thermal photon bunching at microwave frequencies was observed in the cross spectrum of Johnson noise measured with two separate chains of linear amplifiers and square-law detectors \cite{Gabelli04}. Recently, linear amplification followed by quadrature detection was employed to probe microwave photon antibunching of the radiation emitted by a single photon source \cite{Bozyigit10,daSilva10}; the same technique was used to obtain $g^{(2)}(\tau)$ for a continuously pumped single photon source and for thermal and coherent microwaves \cite{Lang11}.  In other work, a protocol was developed to perform state reconstruction of propagating microwave fields from the cross correlations of a dual-path amplification setup \cite{Menzel10}, and researchers have demonstrated state reconstruction of itinerant squeezed \cite{Mallet11} and single photon \cite{Eichler11} microwave fields. In most of these cases, the sought-after signal is overwhelmed by tens of photons of added noise from the microwave preamplifier, and extensive signal averaging and postprocessing of the data are required to extract even lower-order temporal correlations, although the authors of \cite{Mallet11} employed a Josephson paramp to reduce the added noise significantly. By contrast, our Josephson photon counting scheme provides direct access to temporal correlations with a minimal number of preparations of the photon state and minimal signal averaging. To our knowledge, the data presented here constitute the first observation of microwave photon bunching in a coincidence counting experiment.

\begin{figure}[t]
\includegraphics[width=.48\textwidth]{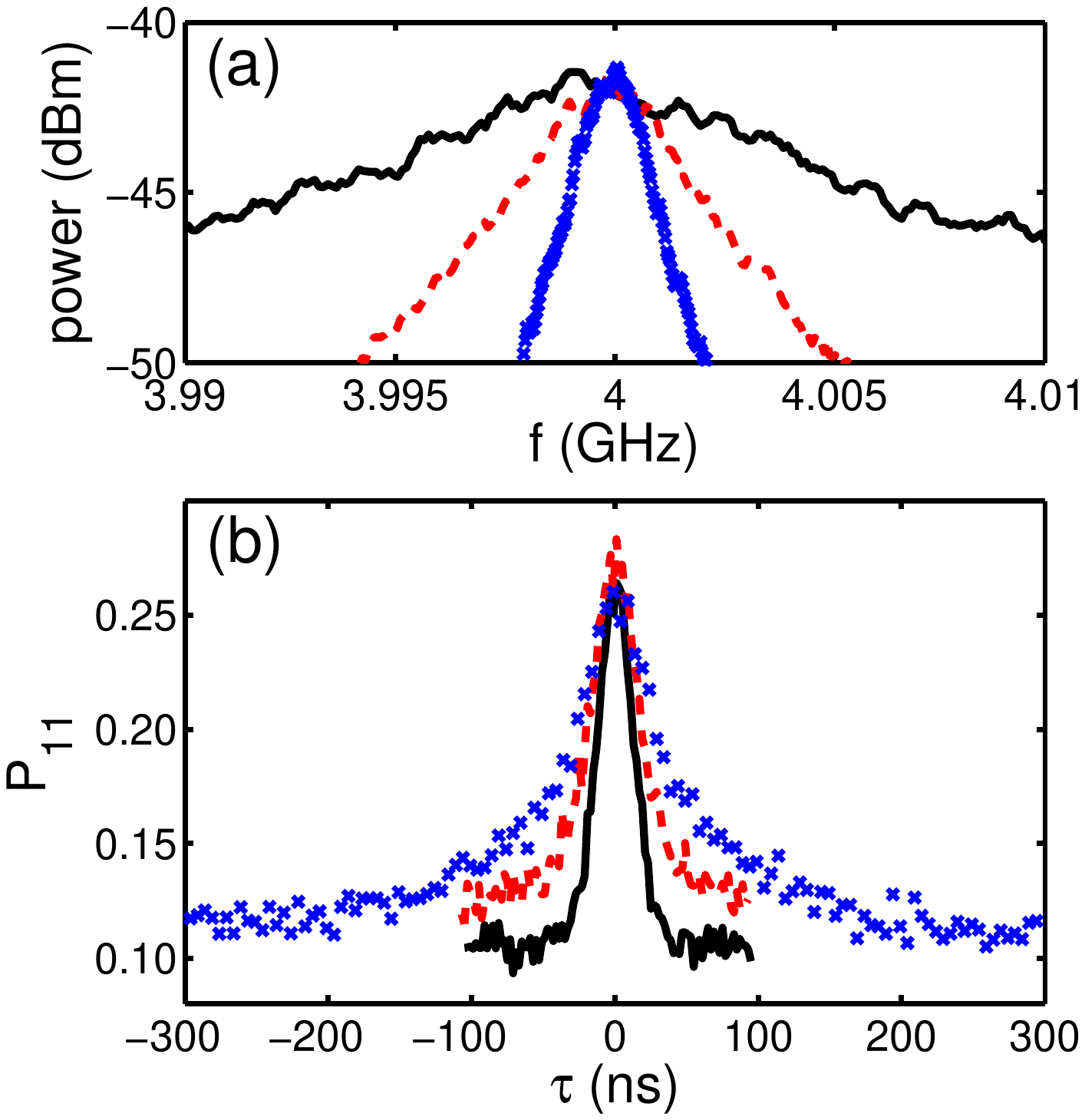}
\vspace*{-0.0in} \caption{Temporal scale of photon bunching. (a) Noise power spectra of thermal sources with bandwidth 2 MHz (cross symbol), 4 MHz (dashed line), and 11 MHz (solid line). (b) The corresponding joint probabilities $P_{11}$ as a function of relative delay $\tau$. The $P_{11}$ enhancement decays on a timescale set by the coherence time of the source, which scales inversely with source bandwidth.}
\label{fig:figure3}\end{figure}

In Fig.~\ref{fig:figure3}, we examine the temporal scale of the photon bunching. Here we have varied the bandwidth of our thermal source by adjusting the low-pass knee of our noise source prior to upconversion. The noise power spectra are plotted in Fig.~\ref{fig:figure3}(a), and the time-domain correlation data are plotted in Fig.~\ref{fig:figure3}(b). We observe that the enhancement of $P_{11}$ decays on a timescale set by the inverse of the noise bandwidth, as expected for a multimode thermal source.

\begin{figure}[t]
\includegraphics[width=.48\textwidth]{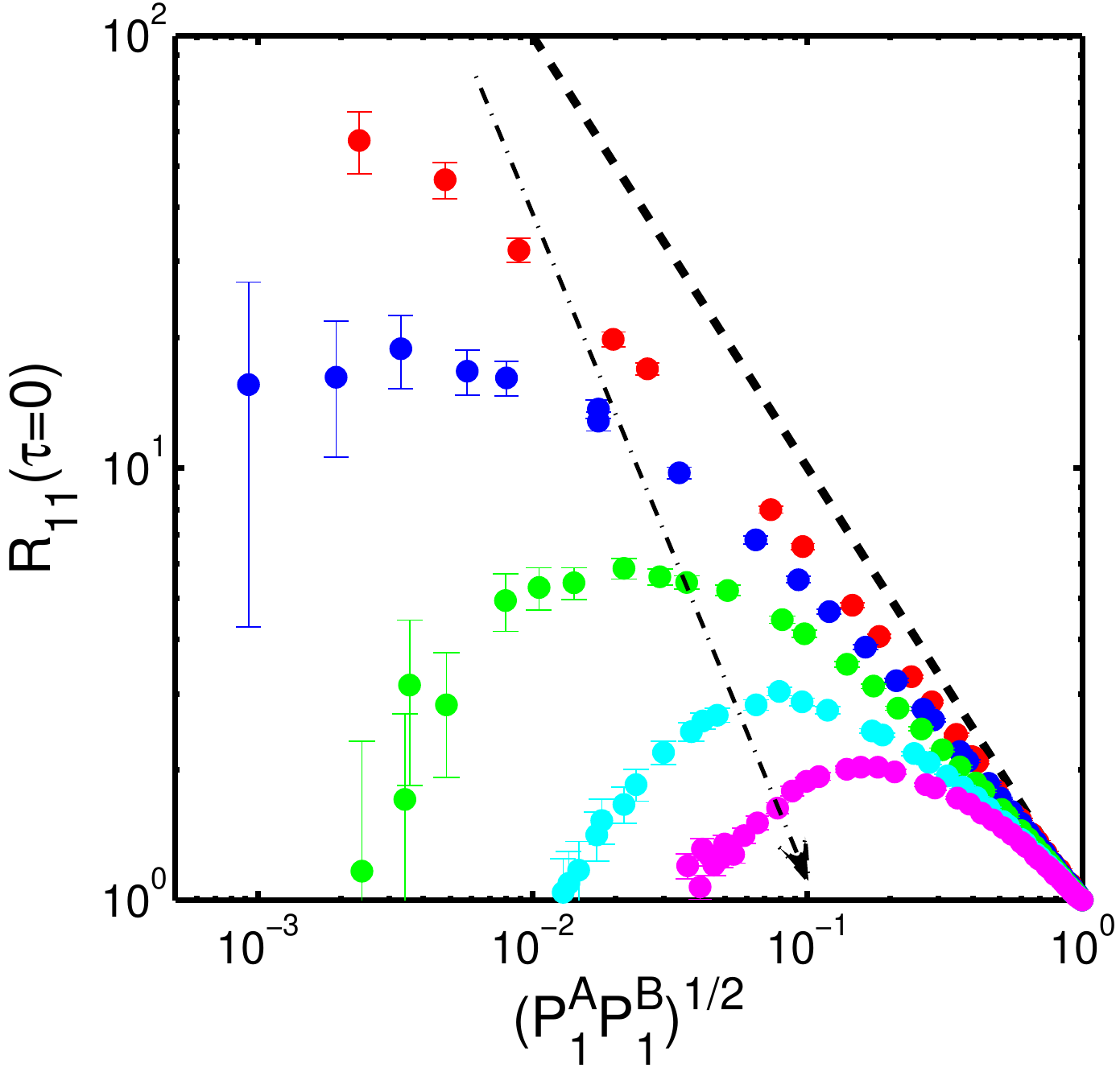}
\vspace*{-0.0in} \caption{$P_{11}$ enhancement $R_{11} \equiv P_{11}/P_1^{\rm A} P_1^{\rm B}$ \textit{versus} $(P_1^{\rm A} P_1^{\rm B})^{1/2}$ for different bias points of the detector; the direction of the arrow indicates shallower bias points, corresponding to increased dark count rates. For each bias point, the probabilities $P_1^{\rm A,B}$ are tuned by adjusting the input microwave noise power. The dashed line corresponds to $R_{11}=1/\sqrt{P_1^{\rm A} P_1^{\rm B}}$. At low power, $R_{11}$ exceeds 2. At lowest input powers, switching is dominated by uncorrelated dark counts and $R_{11}(\tau=0)$ drops to unity. See text for a detailed discussion.}
\label{fig:figure4}\end{figure}

Finally, we have examined the magnitude of the $P_{11}$ enhancement as we varied $P_1^{\rm A,B}$ by tuning the input power of the thermal source. At low powers, we observe an $R_{11}$ at zero relative delay that is much greater than the factor 2 expected from the standard quantum optics treatment of photon counting. In Fig.~\ref{fig:figure4} we plot the zero-delay enhancement factor $R_{11}(\tau=0)$ versus the single detector switching probability $\sqrt{P_1^{\rm A} P_1^{\rm B}}$. For fixed detector bias points, we find that $R_{11}$ scales roughly as $1/\sqrt{P_1^{\rm A} P_1^{\rm B}}$ over a broad range of input power, attaining values as high as 50 at low power. We have performed experimental checks to rule out excess noise from our amplifiers or mixers as the source of the enhanced correlated switching \cite{Chen10}. While we do not have a definitive explanation for the $R_{11}$ enhancement at this time, there are several possibilities. Our strongly coupled Josephson counter interacts coherently with the input microwave field. In a simple model where two detectors are strongly coupled to a single thermal mode, one can show that it is possible to obtain $R_{11}$ far in excess of 2 for interaction times of order half the vacuum Rabi period. We further note that very small deviations from a thermal state can dramatically affect the value of $R_{11}$. For example, if the occupation probability of the $n=2$ Fock state is increased by a small amount $\epsilon$, then in the limit of small $P_1\equiv\sqrt{P_1^{\rm A}P_1^{\rm B}}$, we find $R_{11} = {\epsilon}^{-1}  + \mathcal{O}(P_1)$: counterintuitively, the enhancement of $R_{11}$ is not bounded by 2 but grows inversely with $\epsilon$. We have performed experiments to rule out the possibility of an excess of the $n=2$ Fock state due to downconversion in the detector chip itself. However, we cannot rule out the possibility that our noise source generates photons with statistics that deviate slightly from a thermal state; we plan to investigate this topic in future work.

It should be straightforward to scale this detection scheme to tens of parallelized Josephson junctions to enable number-resolved counting of microwave photons. A multiplexed Josephson photon counter could be used to measure the quantum state of a microwave cavity state in the Fock basis. For a measurement time of 10 ns per junction and a cavity relaxation time of approximately 1~$\mu$s, tens of junctions can interact with the cavity before appreciable decay occurs. Simulations indicate that efficient state reconstruction is possible with a handful of detectors and a small number of coherent displacements of the cavity state. Additionally, a multiplexed Josephson photon counter would allow measurement of higher-order correlation functions of the microwave photon field. Such a tool could be applied to detailed investigations of the statistics of radiation emitted by mesoscopic conductors \cite{Beenakker01}, or to studies of analogue Hawking radiation \cite{Nation09} and the dynamical Casimir effect \cite{Johansson09} in superconducting transmission lines.

\begin{acknowledgments}
We acknowledge inspiring discussions with Luke Govia. This work is supported by the DARPA/MTO QuEST program through a grant from AFOSR.
\end{acknowledgments}

\end{document}